CHAPTER 9
# *Rheology of active fluids*

ALFREDO LANZARO[a], LUIGI GENTILE[b,*]

[a] Institute for Systems Rheology, Guangzhou University, 510006 Guangzhou, China, [b]Department of Chemistry, University of Bari "Aldo Moro", 70126 Bari, Italy

*corresponding email address: luigi.gentile@uniba.it

**Abstract.** This chapter on the rheology of active fluids is an attempt to correlate theoretical and experimental work. A considerable amount of theoretical work and most of the experimental data focus on the rheology of active fluids in a Newtonian matrix, which displays uncommon macroscopic rheological behaviors, such as the apparent superfluid-like state of the pusher suspensions. The failure of the "scallop theorem" for reciprocal swimmers in a non-Newtonian matrix is highlighted. Finally, recent findings concerning the turbulent-like behaviour in concentrated systems are described in detail.

# Contents

## 1.1 Introduction

The most significant parameter in rheology of complex fluids is the Deborah number, $\mathscr{D}$. It is defined as the ratio between the characteristic relaxation time of the fluid, $\lambda$, and the observation time, $t_o$.[1] $\lambda$ represents the characteristic time scale of the material to the applied stress (or deformation). Depending on $t_o$, a flow can therefore be "slow" or "fast" if $\mathscr{D} \ll 1$ or $\mathscr{D} \gg 1$ respectively.

Fluids can generally be divided in:[2] (i) Newtonian fluids, where the relation between the shear stress, $\sigma$, and the shear rate, $\dot{\gamma}$, is linear; (ii) Pseudoplastic (shear thinning) fluids, usually described by the Oswald and de Waele power law equation $\sigma = \sigma_y + K(T)\dot{\gamma}^a$, where $a$, the flow behavior index, is $< 1$, and the yield stress $\sigma_y = 0$, while $K(T)$ is the consistency index at a fixed temperature T; (iii) Dilatant (shear thickening) fluids where the value $a$ of the Oswald and de Waele power law is $> 1$ and $\sigma_y = 0$; (iv) Yield-pseudoplastic and Bingham plastic fluids, which are shear thinning and shear thickening, respectively but both showing a non-vanishing yield stress, i.e., $\sigma_y \neq 0$. In the case of Newtonian fluids, the flow dynamics can be fully described by the Navier-Stokes equations, for which unfortunately an exact solution is not always available due to the complexity of the flow turbulence (high Reynolds number, *Re*), even if various perturbation schemes have been adopted.[3,4]

The relationship between the apparent viscosity $\eta_{app}$ and the volume fraction in suspensions of passive particles is typically described by the Einstein,[5] Batchelor[6] and Krieger-Dougherty[7] equations. On the other hand, a rich rheological phenomenology is observed for lyotropic liquid crystals and polymers when an external force or flow field is imposed. For example, the disturbance arising from the applied flow field can induce a lamellar-to-vesicle transition under shear.[8] Such flow-induced structural changes lead in some cases to regular oscillations of the measured viscosity.[9] This suggests that a further class of materials, characterized by a well-defined time dependence of the measured shear viscosity, should also be considered.

Active fluids can act as shear thinning, shear thickening, shear-unstable fluids or superfluids (vanishing viscosity).[10] The key factors in the rheological response of active fluids are the diffusive transport of the active particles and the disturbance flows induced by the individual particles. The last factor is intrinsically related to the concentration and shape of the active particles. Finally, the medium in which the particles are "swimming" is also of primary importance. In most of the literature the medium is assumed to be a viscous Newtonian fluid. The rheology of active fluids is usually referring to bacteria and algae organisms as main examples along with synthetic designed particles.

### 1.1.1 Low Reynolds number swimming in a Newtonian medium

The motion of fluids surrounding swimming bacteria is modeled by means of the Cauchy equations expressing the conservation of momentum and mass, [11]

$$\rho \left( \frac{\partial \mathbf{u}}{\partial t} + \mathbf{u} \cdot \nabla \mathbf{u} \right) = -\nabla \cdot \sigma - \nabla p + \rho \mathbf{g} \tag{1.1a}$$

$$\nabla \cdot \mathbf{u} = 0 \tag{1.1b}$$

where $t$ is time, $\rho$ is the fluid density, $p$, $\mathbf{u}$ and $\sigma$, are the pressure, the velocity and stress tensor fields respectively, and $\mathbf{g}$ is a body acceleration field due to the presence of external forces (usually gravity). The left hand term of Eq. (1.1a) is often written in terms of the material derivative of the velocity,

$$\frac{D\mathbf{u}}{Dt} = \frac{\partial \mathbf{u}}{\partial t} + \mathbf{u} \cdot \nabla \mathbf{u}, \tag{1.2}$$

which expresses the rate of change of $\mathbf{u}$ as computed from a velocity frame co-moving with the fluid. In order to close the problem, a constitutive equation must be specified to relate $\sigma$ with the strain rate tensor $\mathbf{E} = \frac{1}{2} \left( \nabla \mathbf{u} + \nabla \mathbf{u}^T \right)$. In case of Newtonian fluids, the stress is linear in $\mathbf{E}$, that is,

$$\sigma = -2\eta \mathbf{E}, \tag{1.3}$$

where $\eta$ is the viscosity of the fluid. By substituting Eq. 1.3 into Eq. 1.1a, we obtain the Navier-Stokes equations,

$$\rho \left( \frac{D\mathbf{u}}{Dt} \right) = \eta \nabla^2 \mathbf{u} - \nabla p + \rho \mathbf{g} \tag{1.4a}$$

$$\nabla \cdot \mathbf{u} = 0. \tag{1.4b}$$

Eq. (1.4a) can be made dimensionless as

$$\frac{D\mathbf{u}^*}{Dt^*} = \frac{1}{Re} \nabla^{*2} \mathbf{u}^* - \nabla^* p^* + \frac{1}{Fr^2} \mathbf{g}^* \tag{1.5}$$

where the asterisk indicates a non-dimensional quantity, and $Re = \rho u L/\eta = \rho R^2 \dot{\gamma}/\eta$, and $Fr = u/\sqrt{gL}$, where $L$ is the characteristic lengthscale, $u$ is the particle velocity, $R$ is the particle radius, and $\eta$ is the fluid viscosity that for a suspension of particles is the solvent viscosity $\eta_s$. The Reynolds, $Re$, and Froude, $Fr$, numbers express the ratio between inertial and viscous forces and the ratio between fluid inertia and the strength of the body force, respectively. In the density-matching approach $Fr \rightarrow \infty$ in the case of swimming bacteria with $u \approx 10$ $\mu$m/s, $L \approx 10$ $\mu$m and density close to that of the fluid. Moreover, as the kinematic viscosity of water is $\mu = \eta/\rho \approx 10^{-6}$ m$^2$/s, we also obtain $Re \ll 1$. Therefore, Eq. (1.5) simplifies to the Stokes equation,

$$\mathbf{0} = \nabla^{*2}\mathbf{u}^* - \nabla p^*. \tag{1.6}$$

Eq. (1.6) suggests that the body experiences no net force or torque throughout its motion. Moreover, Stokes equations are linear and independent of time. As a consequence, if the body undergoes a reciprocal motion that does not break the time-reversal symmetry, like the opening and closing of a scallop, it will not experience any net displacement[12–14]. To achieve self-propulsion, swimmers rather need to perform non-reciprocal types of propulsion. Examples include, but are not limited to, wave-like deformations of the body[14] or of the cilia[15,16]. Moreover, the "scallop theorem" is only valid for a single swimmer undergoing inertialess motion through a Newtonian medium. In particular, a violation of the latter condition allows for net displacement also in the case of symmetric swimmers, as will be seen in Section 1.6. For a detailed hydrodynamic description of swimming cells see Chapter **??**.

### 1.1.2 Stokeslet, Stresslet and Rotlet

Stokes equations are linear, implying that any flow can be represented through a combination of simpler flow solutions, such as uniform flows, shear flow, the *Stokeslet* (the flow from a point force), and their derivatives[17]. The derivative of the Stokeslet is the flow from a force dipole. Usually, the force dipole is decomposed into a symmetric component called the **stresslet** and an antisymmetric component called the **rotlet**[17]. The former represents pure strain flow, while the latter is the flow from a point torque. The Stokeslet, stresslet, and rotlet are singularities in the Stokes flow. The full rationalization of the singularities for active particles is provided by Ghose and Adhikari[18] and Saintillan[19]. The rotation of the flagellar bundle of the *Escherichia coli* results in a net thrust $-F_0\mathbf{p}$ with particle orientation $\vec{p}$, which must balance the viscous drag force $F_0\mathbf{p}$ exerted by the cell body as it translates through the fluid. This extensile force dipole results in a net stresslet with magnitude $\zeta \approx -F_0 l$, where $l$ is a characteristic length on the order of the cell size. The negative sign of $\zeta$ indicates a **pusher**, which uses its flagellar tail to push itself through the fluid, while a positive sign would indicate a **puller**. $\zeta$ is often addressed as "activity coefficient" or swimming stresslet. The stresslet describes the single-cell disturbance very well in the far-field,[20] however, near the cell body a rotlet dipole arises due to the equal and opposite torques exerted by the flagellar bundle and cell body as they counter-rotate.[21]

### 1.1.3 Turbulence

To describe active fluids flow, or more properly the flow of active particles in a Newtonian passive fluid, several approaches have been adopted. One approach is to describe the phenomenol-ogy of the active states in terms of fourth- and higher-order expansions of the Navier-Stokes equations[22–24].

$$\nabla \cdot \mathbf{u} = 0 \tag{1.7a}$$

$$\rho \left( \frac{\partial \mathbf{u}}{\partial t} + \lambda_0 \left( \mathbf{u} \cdot \nabla \right) \mathbf{u} \right) = -\nabla p + \lambda_1 \nabla \mathbf{u}^2 - \left( A + C|\mathbf{u}|^2 \right) \mathbf{u} + \eta_0 \nabla^2 \mathbf{u} - \eta_2 \nabla^4 \mathbf{u} + \rho \mathbf{g} \tag{1.7b}$$

The term $\left( A + C|\mathbf{u}|^2 \right) \mathbf{u}$ in Eq. 1.7b represents a quartic Landau velocity potential[25–27] that accounts for the favored polar alignment usually due to the head-tail asymmetries. However, the fact that polar ordering appears only locally but not globally in suspensions of swimming bacteria[28–30] suggests that there are other instability mechanisms[31] that are considered through the $\lambda$ terms. The viscosity coefficients $\eta_0$ and $\eta_2$ are terms of the symmetric and traceless rate-of-strain $\mathbf{E}$ tensor (sometimes called deformation gradient) defined as[23]

$$E_{ij} = \eta_0 \left( \partial_i u_j + \partial_j u_i \right) - \eta_2 \nabla^2 \left( \partial_i u_j + \partial_j u_i \right) + \zeta \left( u_i u_j - \frac{\delta_{ij}}{d} |\mathbf{u}|^2 \right). \tag{1.8}$$

The last term on the right side of Eq. (1.8) is a $d$ independent, $d$-dimensional mean-field approximation of the active nematic stresses due to swimming, in which $\delta_{ij}$ is the Kronecker tensor. The $\zeta$ term is not affecting the linear stability of the model, while it takes into account that an active particle can act as pusher ($\zeta < 0$) like *E. coli*[32–34] or *Bacillus subtilis* or as a puller ($\zeta > 0$) like *Chlamydomonas reinhardtii*.[32,35] Finally, $\lambda_0$ and $\lambda_1$ in Eq. (1.7b) are equal to $\lambda_0 = 1 - \zeta$ and $\lambda_1 = -\zeta/d$. Subsequently, we discuss the following cases:

- For $\eta_0 > 0$ and vanishing $\zeta$, A, C and $\eta_2$, Eq. (1.7b) reduces to the standard Navier-Stokes equation (1.1a) for a passive fluid.

- For $\eta_0 > 0$ and $\eta_2 = 0$, Eq. (1.7b) reduces to an incompressible version of the classical Toner–Tu model.[25–27]

- For $\eta_0 < 0$ and $\eta_2 > 0$, Eq. (1.7b) provides a simplest generic continuum description of turbulent meso-scale instabilities observed in dense bacterial suspensions.[24]

Further details on active turbulence are provided in Chapter **??**.

## 1.2 Rheometry of active fluids and the problem of sensitivity

The rheological response of active fluids is usually non-linear and for pusher ($\zeta < 0$) suspensions, a superfluid-like regime can be observed (Section 1.3). Therefore these bacterial suspensions are usually "low viscosity fluids", with a zero-shear viscosity $\lim_{\dot{\gamma} \to 0} \eta(\dot{\gamma}) = \eta_0 \leq 1$ mPa·s, where $\dot{\gamma}$ is the shear rate[19,34,36]. To characterize low viscosity fluids, conventional (rotational) rheometers need to be equipped with flow geometries that maximize the instrument sensitivity. A short overview of common measuring geometries is provided in this section since in the following paragraphs data coming from several instruments equipped with different geometries are compared. The cone-plate geometry, Fig. (1.1)A, in which the measuring tool is a truncated cone with an angle $\alpha$ below $2°$ for analytical reasons, provides an homogeneous $\dot{\gamma}$ in the gap, and is recommended for suspensions with particle sizes up to approximately $1\,\mu$m. Fig. (1.1)B is the classical coaxial cylinder cell, also known as Taylor-Couette, consisting of an outer cylinder ("cup") of radius $R_1$, a coaxial inner cylinder ('bob') of radius $R_2$, length $L_H$, and a small gap, $H$, between the cylinders, resulting in a simpler flow field. To increase sensitivity a typical choice is the double-wall Couette geometry, which is schematically shown in Fig. (1.1)C. In the double-wall geometry, the sample is loaded in the space between the lower ("cup") and the upper ("bob") tools. The cup is connected to a machine motor, which imposes an angular velocity $\omega$, while the bob is connected to a force transducer which measures the torque $\mathbf{L}$. With reference to the Figure 1.1, the nominal shear rate at the walls $\dot{\gamma}$, is related to $\omega$, by

$$\dot{\gamma} = \omega \left[ \frac{1}{\left(\frac{R_2}{R_1}\right)^2 - 1} + \frac{1}{1 - \left(\frac{R_3}{R_4}\right)^2} \right], \tag{1.9}$$

while L is related to the apparent, averaged shear stress $\overline{\sigma}$ by

$$L = \sigma_1 2\pi R_2^2 L_H + \sigma_2 2\pi R_3^2 L_H = \overline{\sigma} 2\pi L_H \left( R_2^2 + R_3^2 \right), \tag{1.10}$$

with $\sigma_1$ and $\sigma_2$ being the shear stress on the inner and outer bob surface respectively. Finally, the apparent shear viscosity $\eta_{\text{app}}$ is readily obtained as

$$\eta_{\text{app}} = \frac{\overline{\sigma}}{\dot{\gamma}}. \tag{1.11}$$

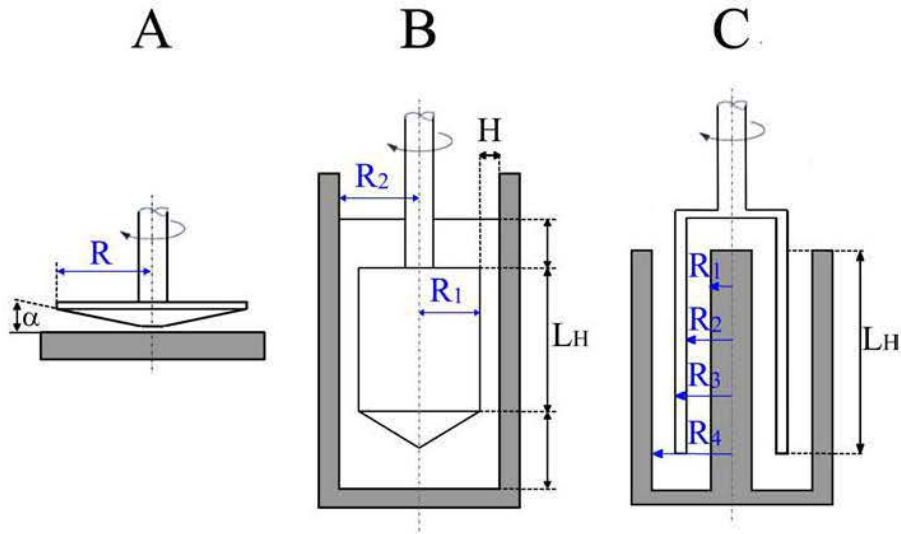

Figure 1.1: Schematic diagrams of (A) cone-plate geometry; (B) Taylor-Couette geometry; (C) double-wall Couette geometry. For all geometries, one part of the tool is fixed and the other rotates around its axis (dashed lines in the figure).

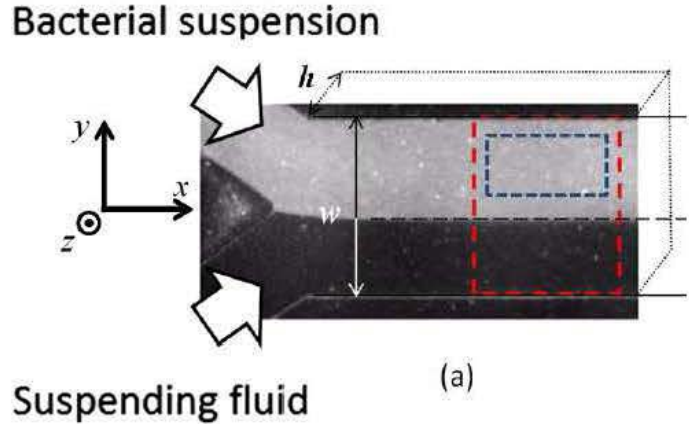

Figure 1.2: The experimental setup used by Gachelin et al.[36] The bacterial suspension is visualised by means of a fluorescent seeding. The flow rate $Q$ of both the bacterial suspension and the buffer are controlled by a syringe pump. The viscosity ratio $\eta_{app}/\eta_s$ is estimated from the ratio between the widths of the two streams $d_1/d_0$ as $\eta_{app}/\eta_s = d_1/d_0$.[37] Reproduced from Ref.[36] with permission from [American Physical Society], Copyright [2013].

The $\eta_{app}$ of pushers suspensions can become as low as 20% of the buffer medium (solvent) viscosity $\eta_s$ over a range of $0.01 \text{ s}^{-1} \leq \dot{\gamma} \leq 1 \text{ s}^{-1}$.[34] Even in case of a very sensitive measurement apparatus such as the ARES-G2 rheometer equipped with a double wall Couette geometry (cylindrical Couette), the corresponding values of the torque $M$ would be $10^{-4} \, \mu\text{N·m} \leq \text{M} \leq 10^{-2} \, \mu \text{ N·m}$, which is below the minimum value measurable by the instrument, $M_{min} = 0.1 \, \mu\text{N·m}$,[38] i.e., leading to inaccurate data. In order to overcome such technical difficulty,[34,39] a Contraves 30 Couette rheometer equipped with a counter-rotation tool, which provides reliable measurements for low viscosity fluids up to shear rates as low as $10^{-6} \text{ s}^{-1}$ can be used[40]. Moreover, Gachelin et al.[36] used a microfluidic Hele-Shaw cell,[37] Fig. 1.2, which allows to determine the $\eta_{app}/\eta_s$ ratio directly from flow visualization experiments, with an error smaller than 10%, thus circumventing the sensitivity issues encountered by rotational rheometers.

## 1.3  The apparent viscosity of active particle suspensions

The analysis of the active fluid dynamics led Hatwalne et al. to predict that the presence of pushers in a Newtonian matrix would lower the bulk viscosity of the suspension with respect to that of buffer, while puller algae would act to increase the viscosity.[41] This was experimentally demonstrated in the work by Sokolov and Aranson focusing on shear viscosity measurements of a pusher species (Bacillus subtilis cells) in a suspension.[42] The shear viscosity of the Bacillus subtilis cells suspension was reduced up to 7 times compared to suspensions of non-motile cells. To explain such an effect, Saintillan proposed a flow curve predictive kinetic model for unbounded dilute suspensions of torque-free particles at low Reynolds number.[43]

The configuration of the suspension is characterized by the particle orientation distribution,

$\psi(\mathbf{p},t)$, where $\mathbf{p}$ is the particle director, which satisfies a Fokker-Planck equation including the effects of the external shear flow, rotational diffusion, and particle tumbling. In this model, it is assumed that the suspension is homogeneous in space, which is a valid approximation in the limit of infinite diluteness, i.e., the inter-particle hydrodynamic interactions are neglected. As a matter of fact, active particle simulations have shown that at low volume fractions swimming particles undergo ballistic motions for most of the time, except for rare particle-particle encounters during which their orientations change, resulting in an effective rotational diffusion over long times, i.e., the single-cell rotational diffusion $d_r$ can be used to model hydrodynamic diffusion.[44–46] The evolution equation for $\psi(\mathbf{p},t)$ is then

$$\frac{\partial \psi}{\partial t} + \nabla_p \left( \dot{\mathbf{p}} \psi \right) - d_r \nabla_p^2 \psi + \frac{1}{\tau_r} \left( \psi - \frac{1}{4\pi} \right) = 0 \qquad (1.12)$$

where $\nabla_p$ is the gradient operator on the surface of the unit sphere, and $\tau_r$ is the correlation time between tumbling events.[44,46] Knowledge of $\psi(\mathbf{p},t)$ allows the direct evaluation of the relative viscosity as expressed by Saintillan[19]

$$\frac{\eta_{app}}{\eta_s} = 1 + \frac{\pi n \ell^3}{6 \ln(2\ell/d)} \left\langle p_x^2 p_y^2 \right\rangle + \frac{3 n k_B T}{\eta_s \dot{\gamma}} \left\langle p_x p_y \right\rangle + \frac{n \zeta}{\eta_s \dot{\gamma}} \left\langle p_x p_y \right\rangle \qquad (1.13)$$

where $\eta_{app}$ denotes the apparent viscosity, i.e., the ratio between instantaneous shear stress and shear rate, $\eta_s$ is the solvent viscosity, $\ell$ and $d$ are the length and the thickness of the rod, $\zeta$ is the strength of the swimming stresslet, $k_B$, is the Boltzmann constant and $T$ is the temperature. The three terms of Eq. (1.13) results from flow-induced, Brownian and active stress, respectively. In the weak-flow limit, i.e., $\dot{\gamma}/d_r \to 0$, where $d_r$ is the rotational diffusion and in the case of smooth swimmers (i.e., tumbling correlation time, $\tau_r \to \infty$) then $\left\langle p_x^2 p_y^2 \right\rangle \to 1/15$ and $\left\langle p_x p_y \right\rangle \to \beta \dot{\gamma}/30 d_r$, where $\beta = \left( (\ell/d)^2 - 1 \right) / \left( (\ell/d)^2 + 1 \right)$ denotes Bretherton's constant which is approximately 1 for a slender swimmer, leading to the low-shear-rate viscosity estimate

$$\lim_{\dot{\gamma} \to 0} \frac{\eta_{app}}{\eta_s} = 1 + \frac{\pi n \ell^3}{30 \ln(2\ell/d)} \left[ \left( \beta + \frac{1}{3} \right) + \left( \beta \frac{\zeta}{k_B T} \right) \right]. \qquad (1.14)$$

Eq. (1.14) expresses solvent viscosity (first term), zero-shear-rate particle contribution to the viscosity in a suspension of passive rods (second term) and the effect of active particles involving the swimming stresslet $\zeta$ (third term). A characteristic time for swimming, $t_c$, i.e., the time for a swimming particle to drag fluid along is length under the effect of its permanent force dipole can be defined[43]

$$t_c = \frac{\pi \eta_s \ell^3}{6 |\zeta| \ln(2\ell/d)}. \qquad (1.15)$$

Fig. 1.3 reports the Saintillan[43] kinetic model at several $d_r t_c$ values, i.e., at several values of the particle size, along with experimental values for pusher and puller suspensions. The model predicts perfectly the experimental data by Lopez *et al.*[34] obtained using a Taylor-Couette geometry for *E. coli* pushers with a bacteria body volume of 1 $\mu m^3$ giving a $d_r t_c$ of 0.42 with $d_r$ = 2.3 rad²/s[47, 48] $\ell$ = 1.9 $\mu m$ and $\zeta$=-7.98·$10^{-19}$ Nm[32] at a volume fraction of 0.11%, while at higher volume fractions a more pronounced decrease in the viscosity is observed and it follows the Carreau law.

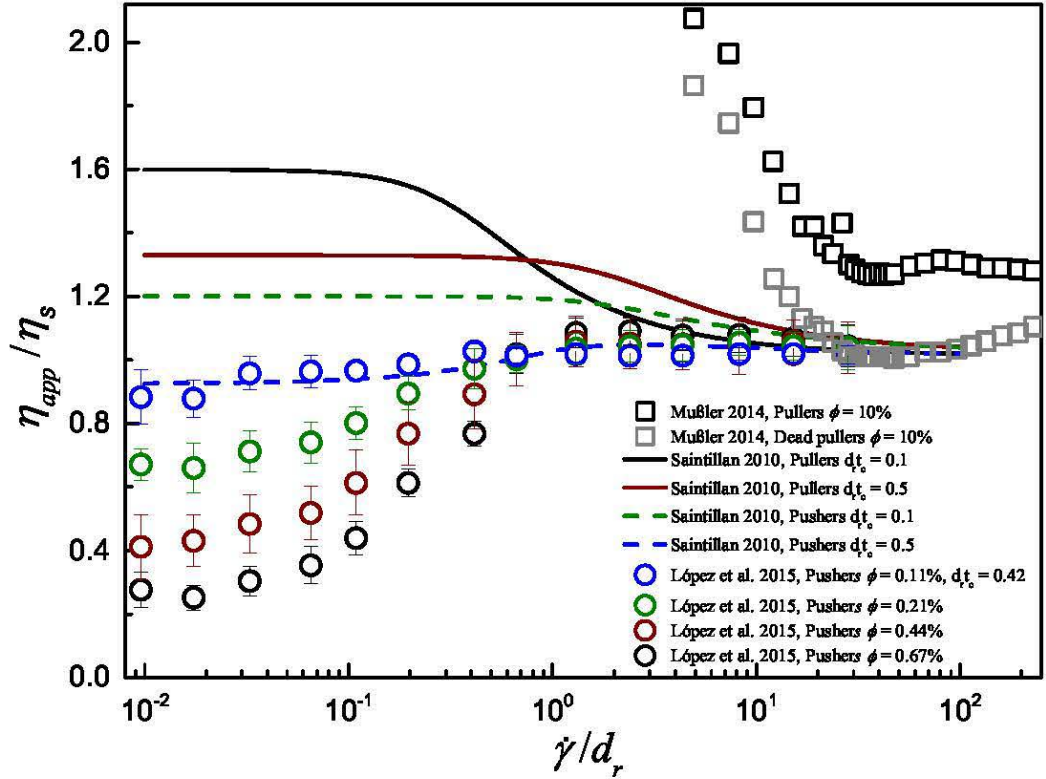

Figure 1.3: Flow curves of suspensions for both pusher and puller swimmers compared with the theoretical prediction by Saintillan[43]. Hollow circles: reduced viscosities of *E. coli* suspensions measured with a Taylor-Couette device at 0.11% $\leq \phi \leq$ 0.65%.[34] Hollow squares: cone-plate measurements of *Chlamydomonas reinhardtii* suspensions showing a decrease in reduced viscosity for the dead (non-motile) pullers.[49] Solid lines: the Saintillan model prediction for dilute suspensions of smooth swimmers as a function of swimming time, $t_c$, and the rotary diffusion, $d_r$, where the Bretherton constant $\beta$ = 1. Theoretical prediction adapted from Ref.[43] with permission from [Springer Nature], Copyright [2010]. Experimental data of pushers adapted from Ref.[34] with permission from [American Physical Society], Copyright [2015]. Experimental data of pullers adapted from Ref.[49] with permission from [the author and supervisor of the thesis], Copyright [2014].

In Fig. 1.3, the relative viscosity of the pusher suspensions display similar trends with volume fraction: (*i*) an active viscous plateau is detectable at very low shear rates, where $\eta_s$ is lower than $\eta_{app}$ (superfluid-like regime); (*ii*) a shear thickening regime appears at intermediate shear rate values bringing the $\eta_{app}$ to a value close to $\eta_s$; (*iii*) a Newtonian plateau at high shear rates where the active contribution to viscosity is negligible. The superfluid-like regime is not a violation of thermodynamic principles, since the bacteria consume chemical energy to balance the viscous dissipation in the flowing suspension, thus allowing for a sustained flow without any applied torque. The Saintillan prediction is made in a very diluted regime, i.e., below the critical volume fraction determined by Martinez et al. et al.[39] as 0.75% for *E. coli*. Above such critical volume fraction, large-scale collective motion emerges from the quiescent state, and the flow becomes nonlinear. Consequently, the apparent viscosity results from a complex interplay between hydrodynamics, particle transport, and hydrodynamic interactions involved in the motion of the suspended active particles as well as the measured gap size (see Section 1.4).[39,50,51] Furthermore, the apparent viscosity is also a function of the bacteria aspect ratio,[52] which is considered to be $\ell/d > 6$[53] for filamentous *E. coli* strains. From an experimental viewpoint, it is worth noting that individual cells of a single species of bacterium vary in length during growth in culture leading to $2.31 < \ell/d < 9.31$.[36] The shear-thinning nature of algae suspensions has been reported by Rafai *et al.*[35] and Mussler *et al.*[54] in particular for a suspension of *Chlamydomonas reinhardtii*, a puller microswimmer. Their main findings confirms Hatwalne *et al.*'s prediction, i.e., the viscosity of the active cells suspensions is higher than the viscosity of the same dead cells (non-motile) suspensions as reported by Sokolov and Aranson.[41,42]

Figure 1.3 reports the flow curves of motile and non-motile *Chlamydomonas reinhardtii* puller suspensions[54] obtained using a cone-plate geometry. The author demonstrates that Taylor-Couette and cone-plate geometries are providing similar values in the investigated shear rate regime. Experimental data are apparently deviating from Saintillan's prediction at low shear rates.[43] However, the volume fraction of 10% is higher than the volume fraction of the pusher suspensions. There is no critical volume fraction for pushers at least based on simulation[55] even though there is an experimental indication that algae interaction plays a role in terms of viscosity enhancement.[56] Finally, it should be highlighted that $d_\mathrm{r}t_\mathrm{c}$ is $\approx 0.03$ considering $d_\mathrm{r}$ = 0.4 rad$^2$/s[32] and $\ell$ = 10 $\mu$m and $\zeta$=4.3·$10^{-17}$ Nm,[57] thus far from the simulated profile at $d_\mathrm{r}t_\mathrm{c} = 0.1$. The shear thinning behaviour is attributed to cell interactions and microstructural changes of the suspension[56,58] or to highly localised viscous heating of the suspension at high volume fraction.[59] Soulies *et al.*[58] have reported the existence of flocs in the shear-thinning regime for *Chlorella* suspensions. Cagney *et al.*[60] investigated flocculation in a suspensions of *T. chuii* at $\phi$ = 5% subject to several shear rates. The shear thinning behaviour at low shear rates has been attributed to the detected flocs at 10 $s^{-1}$, whereas no flocculation was reported at higher shear rates. Cagney *et al.*[60] also reported shear-thinning behaviour for the *Phaeodactylum tricornutum* suspension at $\phi$ = 10%, even though Wileman *et al.*[61] reported a Newtonian behaviour between $\phi$ = 0.5 and 8%. They suggested the presence of a critical volume fraction at which suspensions of *P. tricornutum* become non-Newtonian, as has been observed for some *Chlorella* species.[58,61,62] *Chlorella* species of the green algae genus *Chlorella* have shown disparate behaviour. Wu and Shi[62] studied *Chlorella pyrenoidosa* and observed Newtonian behaviour up to $\phi$ = 15%, while above this concentration the viscosity increased dramatically and could not be described by Einstein's equation, Eq. (1.16). Moreover, at $\phi > 17.5\%$ a yield stress behaviour was observed and described by the Herschel-Bulkley model. Zhang *et al.*[63] examined suspensions of freshwater and marine *Chlorella sp.* reporting shear-thinning behaviour at all volume fractions ranging from 4 to 8%. On the other hand, Wileman *et al.*[61] reported Newtonian suspensions of *Chorella vulgaris* for $\phi < 2\%$.

### 1.3.1 Concentration dependence of the apparent viscosity

Fig. 1.4 reports the $\eta_{app}/\eta_s$ from 0.1 to 13% and from 5 to 18% for pushers and pullers, respectively. Accurate measurements of pushers dispersions (*E.coli* strains) have been obtained by Lopez *et al.* and Martinez *et al.* using a Taylor–Couette geometry. Other measurements have been obtained by using microfluidic rheometers [36,64] and are in agreement with the ones reported here. The generalized Einstein's equation could be adopted in the diluted regime of pusher particles, even though it was originally developed to describe dilute regime ($\phi < 5\%$) of solid spherical particles,

$$\eta_{app} = \eta_s \left(1 + [\eta]\,\phi\right), \tag{1.16}$$

where $[\eta]$ is the intrinsic, dimensionless viscosity equal to 5/2 for solid spherical particles, $\eta_s$ is the solvent viscosity, i.e., the viscosity of the suspending medium. However, here $[\eta] = K\left(\tau_r/t_c\right)$ [34,65,66], in which $\tau_r$ characterizes the directional persistence of a swimming trajectory, $\tau_r = 1/2d_r$ [48], while $K \propto \left(A - B(\tau_r/t_c)\right)$ where A and B depend on the bacterium shape [66–68]. In this diluted regime the relative viscosity is linearly decreasing with the volume fraction for shear rate between 0.022 and 0.075 s$^{-1}$ (Fig. 1.4A) following Eq. (1.16). Lopez *et al.* reported $[\eta] \approx -120$ for ATCC9637 strain with low content of $O_2$ (empty circles) and $[\eta] \approx$ -200 for ATCC9637 strain with high content of $O_2$ and L-serine and (grey circles) and $[\eta] \approx$ -260 for RP437 strain with high content of $O_2$ and L-serine (black circles), using Martinez *et al.* [39] data $[\eta] \approx$ -150 for AB1157 strain with a motility buffer. The intrinsic viscosity is a function of both the culture strain and the mobility buffer. In all cases, the active plateau viscosity becomes independent of the volume fraction from 0.75% up to 2.4%. Moreover, in the presence of $O_2$ and L-serine the local viscous dissipation is macroscopically compensated by the swimming activity, i.e., the viscous response reaches zero.

Mussler *et al.* [54] reported accurate viscosity measurements on *Chlamydomonas reinhardtii* puller dispersions using a cone-plate geometry in a wide range of volume fraction, Fig. 1.4B adopting the very well-known Krieger-Dougherty model [59]

$$\eta_{app} = \eta_s \left(1 - \frac{\phi}{\phi_{max}}\right)^{(-[\eta]\phi_{max})} \tag{1.17}$$

where $\phi_{max}$ is the maximum packing volume fraction, which is approximately 0.64 for random close packing [69]. However, the Krieger-Dougherty equation might fail to describe data at the highest volume fraction. Cagney *et al.* [60] demonstrates that the degree to which the volume fraction of Pullers affects the rheology is dependent on the strain rate, at low strain rates the rheology is strongly dependent on the cell volume fraction. This evidence suggests that there is a change in the physical mechanism by which the suspended cells affect the fluid rheology. Cagney *et al.* [60] suggested that this is related to the appearance of flocs at low shear rates, since the tendency to form flocs induces a significant increase in the observed viscosity.

## 1.4 Concentrated Systems and the Onset of Turbulence

One of the most important features of the suspension rheology of pushers is the transition observed at volume fractions $\phi$ above $\phi_c = 0.75\%$. Above $\phi_c$, the system becomes a "super-fluid",

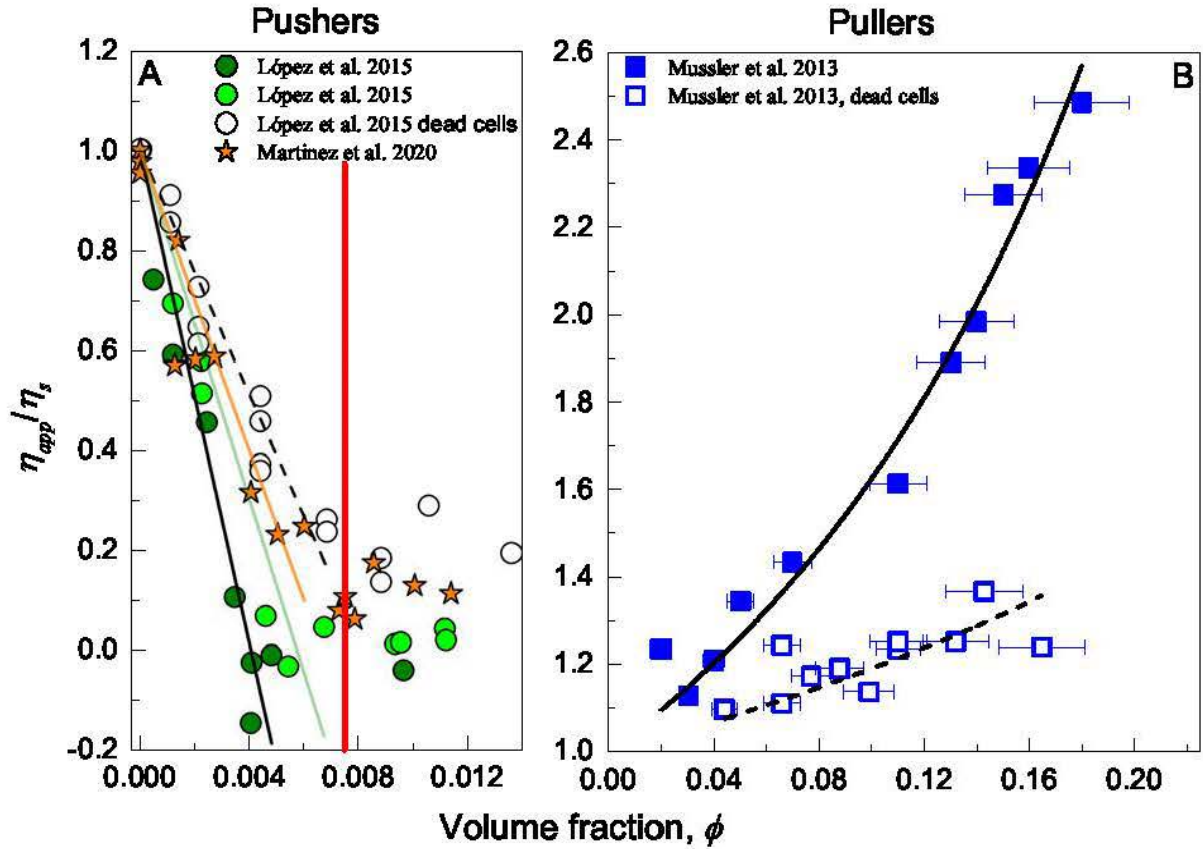

Figure 1.4: Relative viscosity as a function of volume fraction for pushers (A) and pullers (B). A: measurements of *E. coli* suspensions performed with a Taylor–Couette device showing that $\eta_{\mathrm{app}}/\eta_s$ decays towards zero with increasing volume fraction. The solid lines in panel A are linear fittings. B: relative viscosity of *Chlamydomonas reinhardtii* suspensions measured in a cone-plate geometry as a function of the volume fraction.[54] The solid lines in the panel B are Krieger-Dougherty fittings, Eq. 1.16, in which $\phi_{\max} = 0.63$ and 0.64 and $[\eta] = 4.46$ and 1.60 for motile (alive cells) and non-motile (dead cells), respectively. Experimental data of pushers adapted from Ref.[34] with permission from [American Physical Society], Copyright [2015]. Experimental data of pushers adapted from Ref.[39] public data. Experimental data of pullers adapted from Ref.[54] with permission from [IOP Publishing], Copyright [2013].

i.e., the measured macroscopic shear viscosity becomes negligible, and the flow field becomes non-linear.[39]

Guo *et al.*[70] studied the velocity field of *E. Coli* suspensions under different shear rate amplitudes $\dot{\gamma}_0$ through a planar oscillatory shear device consisting of two plates separated by a gap, $H$, i.e., applying an oscillatory shear flow. The key result is that, under weak shear conditions, the velocity gradient concentrates within the bulk of the flow, so that a boundary layer $h_s$ near the top and the bottom of the flow geometry is formed, where the velocity gradient is negligible. For volume fractions lower than 0.01 approximately, the non-linearity of the velocity profile disappears (see Fig. 1.5).

The authors extracted the enstrophy of the flow due to bacterial swarming from the velocity measurements,[71] $\Omega_y = \langle \omega_y/2 \rangle$ where $\partial_x u_z - \partial_z u_x$ is the in-plane vorticity. Interestingly, the measurements of $h_s$ vs. $\Omega_y/\dot\gamma_0$ conducted at various $H$ overlapped on the same curve. Based on a previous model of active fluids suggesting the non-monotonicity of the shear rate-stress curve,[72] the authors proposed a phenomenological model for $h_s$ dependence on $\dot\gamma_0$,

$$\frac{h_s}{H} = \begin{cases} \frac{1}{2}\left(1 + \frac{\dot\gamma_0}{C\sqrt{2\Omega_y}}\right) & \text{if } \dot\gamma_0/\sqrt{\Omega_y} \leq C\sqrt{2} \\ 1 & \text{if } \dot\gamma_0/\sqrt{\Omega_y} \geq C\sqrt{2} \end{cases} \tag{1.18}$$

where $C$ is a constant of order unity. The equation is in good agreement with the experimental observation. From the measurements of $\mathbf{u}$, the velocity-velocity spatial correlation is defined as

$$C_v(r) = \frac{\int\int d\mathbf{r}_i d\mathbf{r}_j\,(\mathbf{u}(\mathbf{r}_i))\,\delta\,(r_{ij}-r)}{\int d\mathbf{r}_i\,(\mathbf{u}(\mathbf{r}_i)\cdot\mathbf{u}(\mathbf{r}_i))}, \tag{1.19}$$

where $\mathbf{r}$ is the position vector, and the correlation length $\ell_c$ is defined as the distance at which $C_v$ becomes equal to $1/e$. The $\ell_c$ varies approximately linearly with the gap height for small values of $H$ before leveling up to a plateau, but does not change very much with bacterial concentration. The results by Guo *et al.*[70] clearly indicate the importance of confinement in modulating the super-fluidity of pusher solutions.

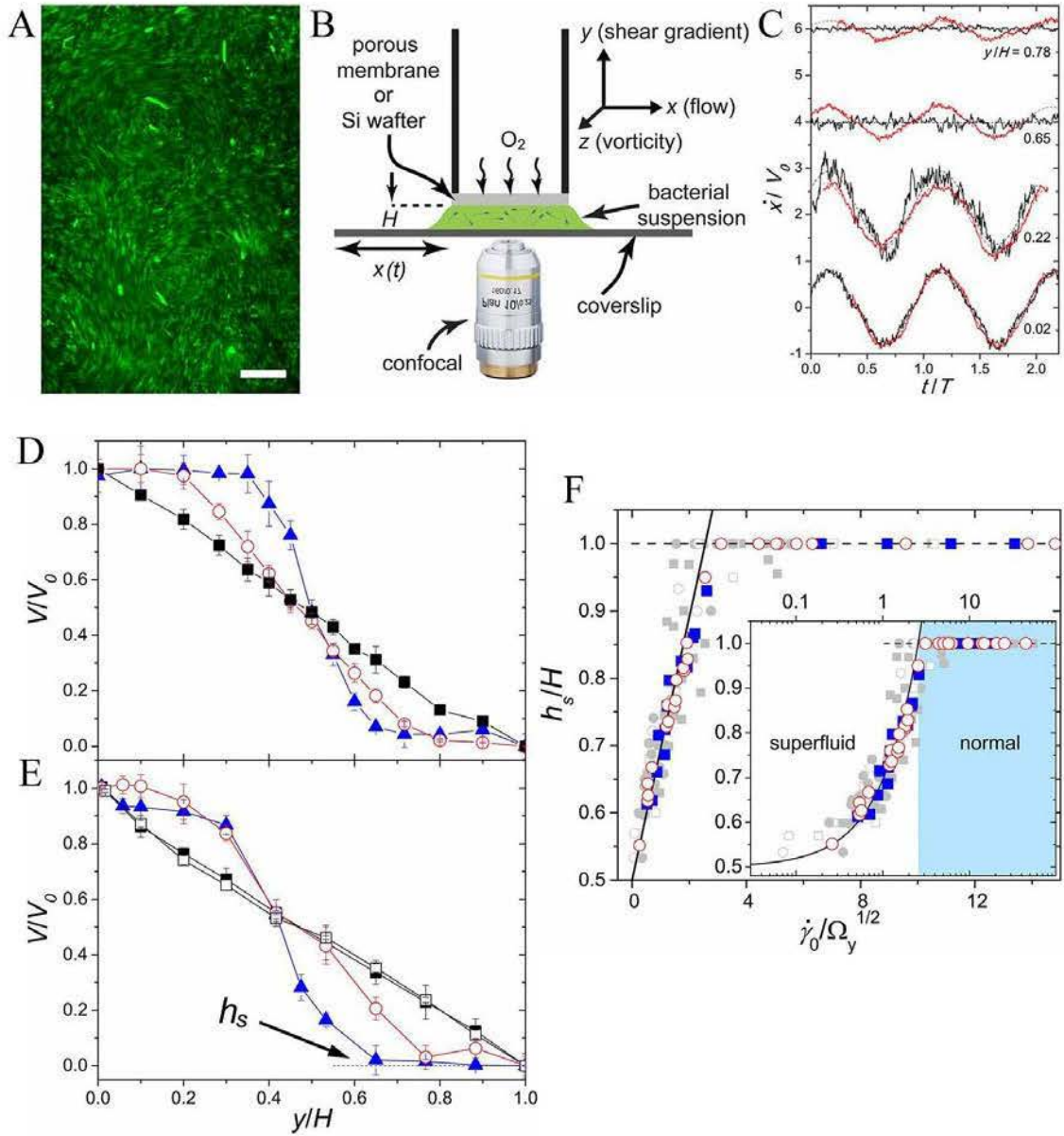

Figure 1.5: Shear banding due to bacterial swarming: (A) an example of bacterial swarming; (B) the experimental setup for flow visualization; (C) temporal variation of velocity for different $y/H$; (D) effect of shear rate on the velocity profile for a fixed bacterial concentration $n = 50 n_0$ and $\dot{\gamma}_0 = 0.42$ s$^{-1}$ (black squares), 0.16 s$^{-1}$ (red circles) and 0.055 s$^{-1}$ (blue triangles). (E) Normalised shear profiles at $\dot{\gamma}_0 = 0.16$ s$^{-1}$ and $n = 10 n_0$ (black squares), $n = 40 n_0$ (red circles) and $n = 50 n_0$ (blue triangles), with $n_0 \approx 8 \cdot 10^8$ mL$^{-1}$. Fig. 1.5F: the width of the non-dimensionalised boundary layer $h_s/H$ plotted vs $\dot{\gamma}_0/\Omega_y$ to display the universality of the shear banding behaviour. Adapted from Ref.[70] with permission from [National Academy of Sciences], Copyright [2018].

Martinez *et al.*[39] visualized the flow of several *E. Coli* suspensions through a cone and plate geometry at a fixed $H \approx 170$ $\mu m$ (see Fig.1.1B). The authors compared their results with a prediction for the critical volume fraction $\phi_c(H,u,\tau_r)$ for the onset of unstable flow based on continuum kinetic theory,[73,74]

$$\frac{\phi_c(H,u,\tau_r)}{\phi_c^\infty(400,15,\tau_r)} \approx 1 + \frac{3}{10}\left(2\pi\frac{u\tau_r}{H}\right) + \frac{1}{5}\left(2\pi\frac{u\tau_r}{H}\right)^2, \tag{1.20}$$

where the asymptotic value $\phi_c^\infty(400,15,\tau_r)$, is the critical volume fraction calculated at gap size of 400 and *E. coli* swimming speed, $u$, of 15 $\mu m/s$, $\phi_c(H,u,\tau_r) \approx 0.75\%$, and $\tau_r$ is the characteristic time between two tumbling events. Interestingly, Eq. 1.20 successfully predicts the onset of the measured unstable flow. At the crossing of $\phi_c(H,u,\tau_r)$, the velocity vs $H$ profiles evolve from linear to non-linear while the measured shear viscosity vanished within this range of $\phi$ (see Fig. 1.6). Similarly to the work of Guo *et al.*,[70] the super-fluidity regime is accompanied by the formation of large vortex structures, which result in the correlation length, $\ell_c$, to suddenly increase from near-zero to $>100$ $\mu m$ at the crossing of $\phi_c(H,u,\tau_r)$.

## 1.5   Rheology of active polar liquid crystals

The rheological behaviour of active polar liquid crystals was predicted by Negro *et al.*[75] by extensive lattice Boltzmann simulations. The system sets into a lamellar phase, i.e., the lyotropic liquid crystalline phase, for the symmetric composition of an active polar component and a Newtonian passive fluid. The authors investigated the behaviour of the relative viscosity of a polar liquid crystal phase as a function of the activity coefficient. The authors adopted the

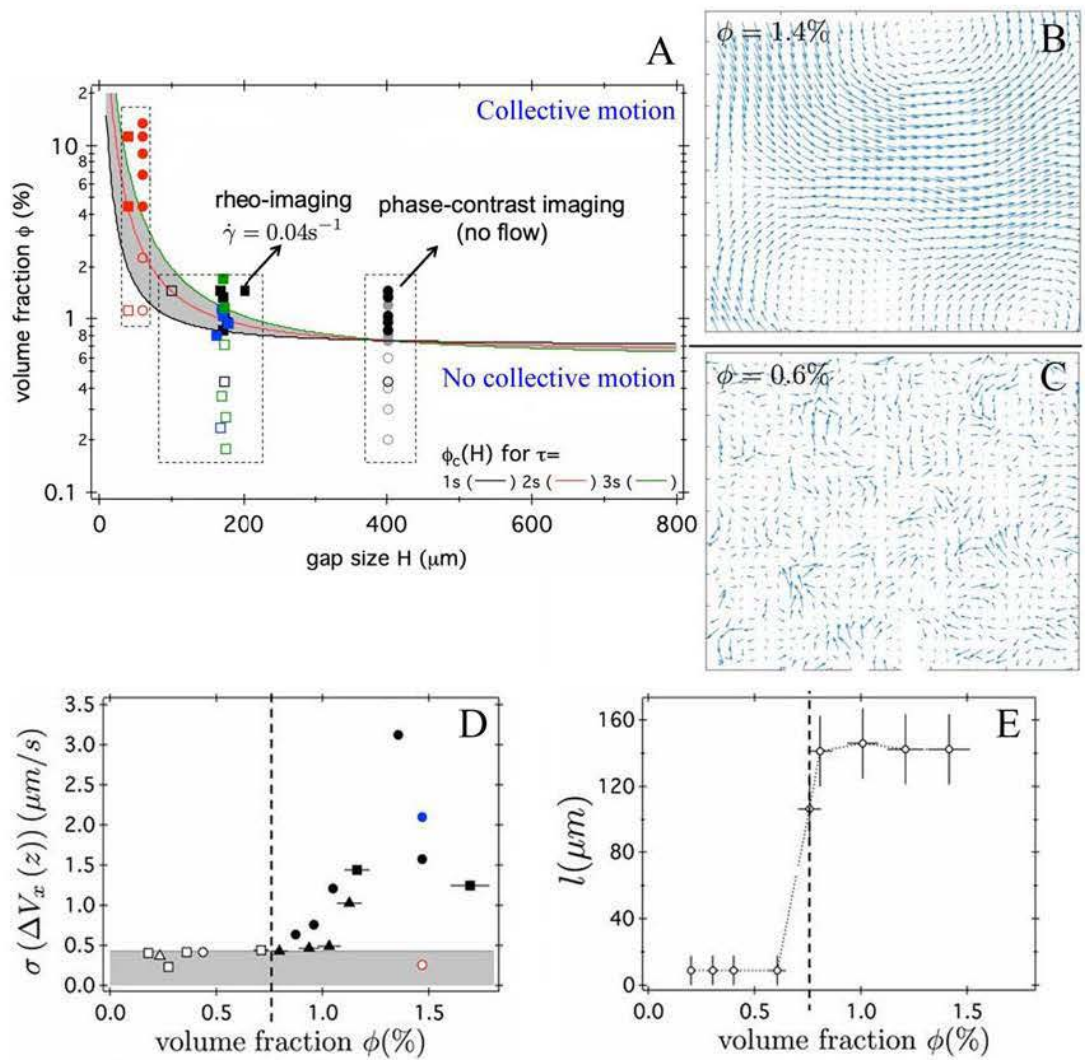

Figure 1.6: Effects of pushers volume fraction $\phi$ and gap size $H$ on the observed flow field.[39] (A): the theoretical prediction based on continuum kinetic theory[73] successfully predicts the transition from individual to collective motion in the $\phi - H$ phase diagram. (B) and (C): examples of the measured velocity field in the individual (A) and collective (B) flow regimes. (D) and (E): the standard deviation $\sigma$, and the correlation length $\ell_c$ of the velocity field are greatly increased for $\phi \geq \phi_c = 0.75\%$. Adapted image courtesy of Martinez et al.[39]

Eriksen number, $Er$, and the active Eriksen number, $Er_a$ as relevant nondimensional quantities to compare external shear flow and active forces in the system.

$$Er = \frac{\eta_s \dot{\gamma}}{B}, \qquad (1.21)$$

$$Er_a = \frac{|\zeta|}{B}, \qquad (1.22)$$

where $B$ is the lamellar compression modulus and the $\zeta$ activity coefficient. In the shear-free system a transition from the lamellar phase to a phase with moving active droplets is found at $Er_a \approx 0.11$ as predicted by Bonelli *et al.*[76] For $Er_a > 1.1$ the system entered into a mixed phase, characterized by chaotic velocity patterns.[77] Fig. 1.7A shows a decrease in the viscosity with increasing activity. The authors demonstrated that the lamellar phase region is extended towards larger $Er_a$ with respect to the shear-free system due to the tendency of the lamellae to align with the flow 1.7B.

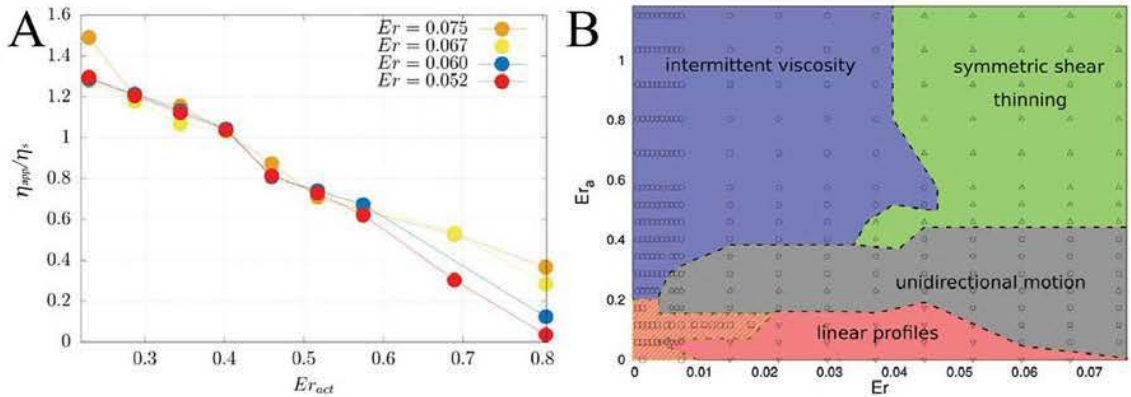

Figure 1.7: Relative viscosity of active polar emulsion varying the active Ericksen number $Er_a$ (A). Dynamic phase diagram in terms of $Er$–$Er_a$ (B). Adapted from Ref.[75] with permission from [RSC], Copyright [2019].

## 1.6 Matrix viscoelasticity affects the swimming behaviour

The matrix where active particles are situated plays a crucial role in the observed swimming behaviour. Gastric and cervical fluid, for example, are viscoelastic fluids, and they are both inhabited by microswimmers.[78,79] Blood also exhibits viscoelastic properties such as an extensional relaxation time on the order of a few ms.[80] Even though blood is normally sterile, it can potentially be infected by *E. Coli* and by Gram positive bacterial swimmers.[81] Matrix viscoelasticity can alter significantly the behaviour of swimmers with respect to the Newtonian case. The majority of matrices where swimmers are found can be described as polymer solutions. Such kind of fluids are characterised by a relaxation time $\lambda$ ranging from milliseconds (in the dilute regime) up to a few seconds or above in the concentrated regime.[82–84] Apart from the Reynolds number, the motion of an actively swimming particle will therefore also be affected by the Deborah number (see Section 1.1), which in this special case is defined as[85]

$$\mathscr{D} = \lambda \omega, \tag{1.23}$$

where $\omega = 1/t_o$, is the frequency of flagella rotation. As $\omega$ of the swimming bacteria is typically on the order of 20-50 Hz, it is clear that $\mathscr{D}$ can in many cases exceed unity, leading to a non-equilibrium configuration of the microstructure of the matrix. This has the important consequence that even reciprocal swimming can induce propulsion into a viscoelastic matrix, which constitutes an apparent violation of the "scallop theorem". This was observed first by Lauga,[86] who estimated that the time-averaged speed $\langle U \rangle$ of a swimmer with radius $a$ in a viscoelastic medium descried by the Oldroyd-B model[87] is given by

$$\langle U \rangle = Ka\omega \frac{\mathscr{D}}{1 + \mathscr{D}^2} \left( 1 - \frac{\eta_s}{\eta} \right), \tag{1.24}$$

where $a$ is the particle radius, $K \approx 2$ and $\eta_s$ and $\eta$ are the solvent and the polymer contribution to the overall matrix viscosity. In the limit of $\mathscr{D} \ll 1$, $\langle U \rangle \to 0$ and the "scallop theorem" is recovered.

The observation made by Lauga motivated several works focusing on the motion of reciprocal swimmers into viscoelastic media. Keim et al.[88](see Fig. 1.8) demonstrated that asymmetric dimers immersed in a Boger fluid (a constant viscosity, yet elastic) medium are displaced when driven by a time-symmetric magnetic field. The same motion was not observed into a Newtonian medium. The propulsion velocity of the swimmers appeared to depend on the frequency, but not on the amplitude of the applied magnetic field. Qiu et al.[89] investigated the behaviour of magnetically-activated "micro-scallops" in either shear thinning or shear thickening suspending media. By modulating the duration of opening and closing phases of the scallop, the authors were able to achieve net positive displacement in both the non-Newtonian matrices.

The precise mechanisms which leads matrix viscoelasticity to alter the swimming of active particles are currently under debate. It has been suggested[47] that the first normal stress difference $N_1$ observed for polymer solutions and melts in the perpendicular direction with respect to that of shear[85] affects the swimming of cilia and other motor organs of bacteria. Moreover, because the swimming motion curves the streamlines around the bacteria, the flow can also become locally unstable, thus triggering modifications of the particle motion with respect to the Newtonian case. Such transition occurs when the Pakdel-McKinley condition,[90,91]

$$\left[ \frac{\lambda_1 \langle u \rangle}{\mathfrak{R}} \frac{\sigma_{11}}{\eta_0 \dot{\gamma}} \right]^{\frac{1}{2}} \geq M_{\text{crit}}, \tag{1.25}$$

is met. In Eq. 1.25, $\eta_0$ is the zero-shear viscosity, $\langle u \rangle$ is the average speed of fluid, $\mathfrak{R}$ is the radius of curvature of the streamline, $\sigma_{11}$ is the stress in the streamline direction and $M_{\text{crit}}$ is a critical parameter depending on the specific rheological properties of the fluid.

Patteson et al.[47] studied the influence of viscoelasticity of the medium on the swimming and tumbling of E. Coli. They used three different polymers, and tuned polymer concentration as well as molecular weight. They found that increasing polymer concentration results in an increase of the bacterium velocity and in a reduction of the tumbling frequency. In order to explain the decrease of tumbling induced by polymers, the authors theorized that the shearing of the polymer solutions produces a normal force $N_1$ due to polymer chains being stretched which interferes with the cilia motion therefore inducing a decrease of wobbling. In order to prove this, a minute quantity of tagged DNA was introduced into the solution. The conformation distribution changed with the shear generated by the bacterium, therefore confirming that polymer deformation plays an important role in bacterial swimming.

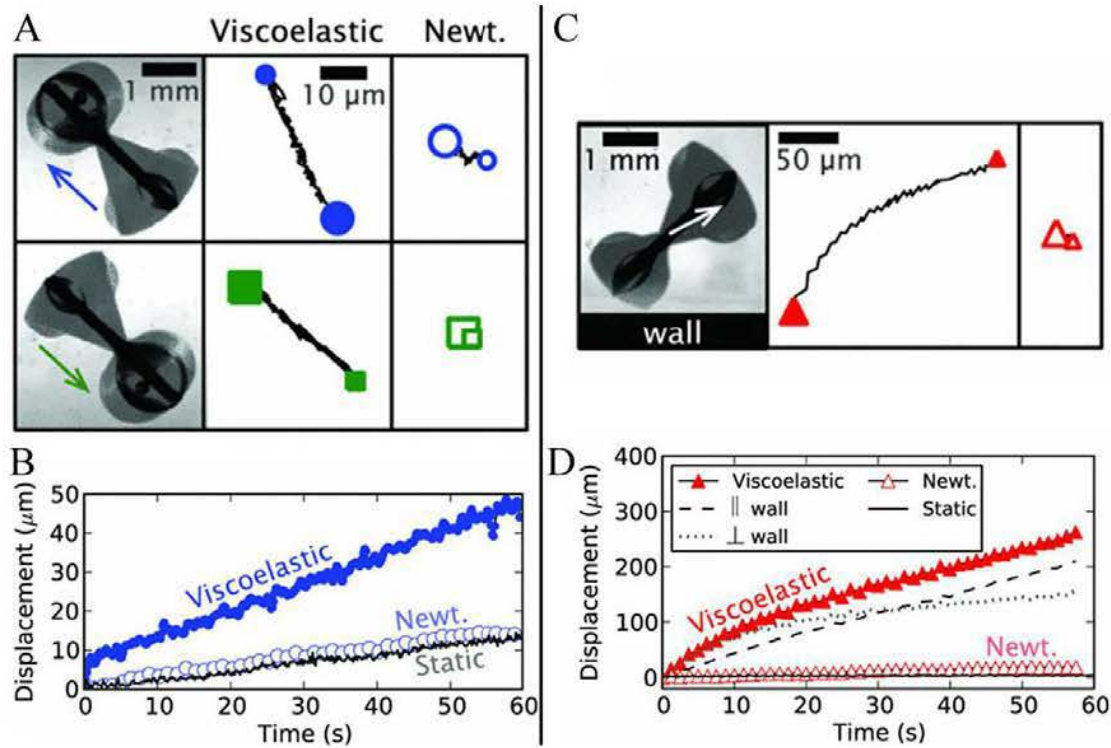

Figure 1.8: The effects of matrix viscoelasticity on the motion of synthetic swimmers. (A) and (B): adding a small quantity of an high molecular weight polymer induces a net displacement of an asymmetric swimmer at $\mathscr{D} \approx 6$, while no displacement is observed in a Newtonian matrix. (C) and (D): the effects of matrix viscoelasticity on the displacement of an active, synthetic swimmer located near the wall. Adapted from Ref.[88] with permission from [AIP Publishing LLC], Copyright [2012].

Gagnon *et al.*[92] studied the flow of magnetically-driven bead-rod particles into a wormlike micellar solution for $0.29 \leq \mathscr{D} \leq 29$. At the lowest $\mathscr{D}$, the particles exhibited a small displacement in the direction of the magnetic field. Such displacement increased at $\mathscr{D}=2.9$, and disappeared at the largest $\mathscr{D}$. In order to assess such result, the experiments were coupled with birefringence and velocity measurements to measure local stress and velocity profiles. At intermediate $\mathscr{D}$, striation regions with different values of birefringence were observed near the rod. This indicates the presence of a stress field which caused the observed particle displacement. Rheometric experiments in oscillatory flow demonstrated that the amplitude of the first normal stress difference of the micellar solution at a fixed strain reduced after a few cycles. Because of this, it was proposed that the flow of the asymmetric bead-rod system was driven by a local imbalance of the first normal stress difference $N_1$.

Han *et al.*[93] investigated the swimming behaviour of micro-scallops made by a variable number of building blocks into a shear thinning fluid. Surprisingly, it was found that increasing the size of the swimming units from 4 to 10 cubes resulted into a reversal of swimming direction, which was attributed to a variation of the local viscosity field encountered by the swimming objects.

Gomez-Solano *et al.*[94] studied the dynamics of Janus particles (i. e., synthetic colloids with artificially tuned anisotropic interactions[95]) in a viscoelastic polymer solution. By imputing a laser beam into the system and thereby moving the particles, they observed a dramatic increase of the rotational diffusion, which was related to the viscoelasticity of the medium as well as to the achieved $\mathscr{D} > 1$.

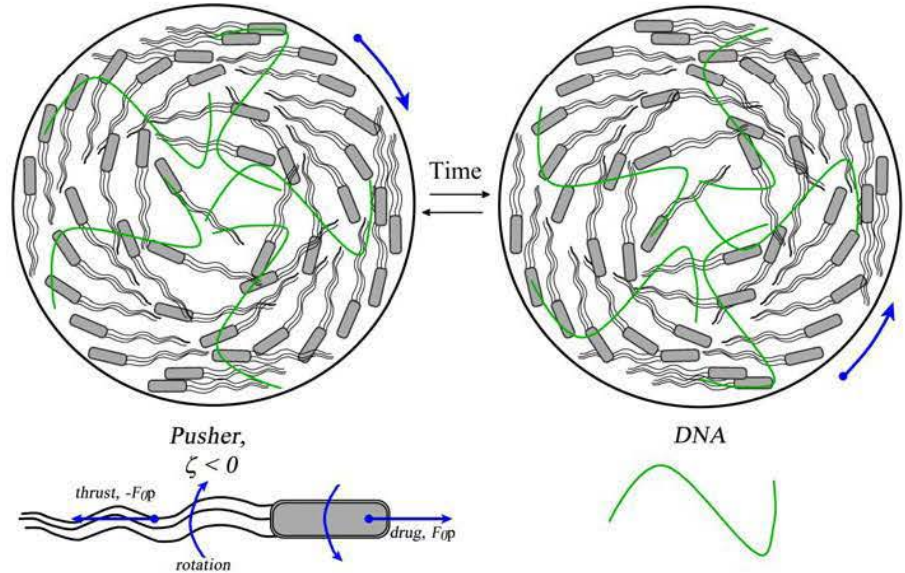

Figure 1.9: Schematic representation of the dramatic effect of adding a small amount of DNA on the flow field of swimming bacterial suspensions. The presence of the DNA macromolecule induces the onset of a sample-spanning rotational motion. The direction of the flow field changes from clockwise to anti-clockwise over a defined time period[96].

The recent results obtained for active fluids in turbulent-like regime can be critically compared with previous investigations focusing on polymer solutions. Grosiman and Steinberg[97,98] discovered that dilute solutions of high molecular weight polymers give rise to a sudden increase of the measured shear viscosity, as well as to turbulent-like features flow effects even at $Re \ll 1$ in either von Karman swirling disks or microfluidic curvilinear geometries, provided that the Weissenberg number $Wi = \lambda_1 \dot{\gamma}$ is of order unity. Such Elastic Turbulence (E.T.) regime displays unique properties different from that of classical (inertial) turbulence.

Most remarkably, in E. T. the formation of a layer with enhanced magnitude of both local strain rate and velocity fluctuations near the boundaries of the flow was observed.[99] The width of such layer is $Wi$-independent, and varies only very weakly with the polymer concentration.[100] The maximum measured local strain rate within the boundary layer tends to saturate for sufficiently high values of $Wi$, which was related to a sudden coil-stretch transition of the polymer chains within the boundary layer, occurring at $Wi \approx 1$.[99] Therefore, the key features of the turbulent-like regime observed for active fluids (i.e., vanishing viscosity and appearance of a boundary layer with vanishing strain rate) are *opposite* with respect to what is observed in E. T..

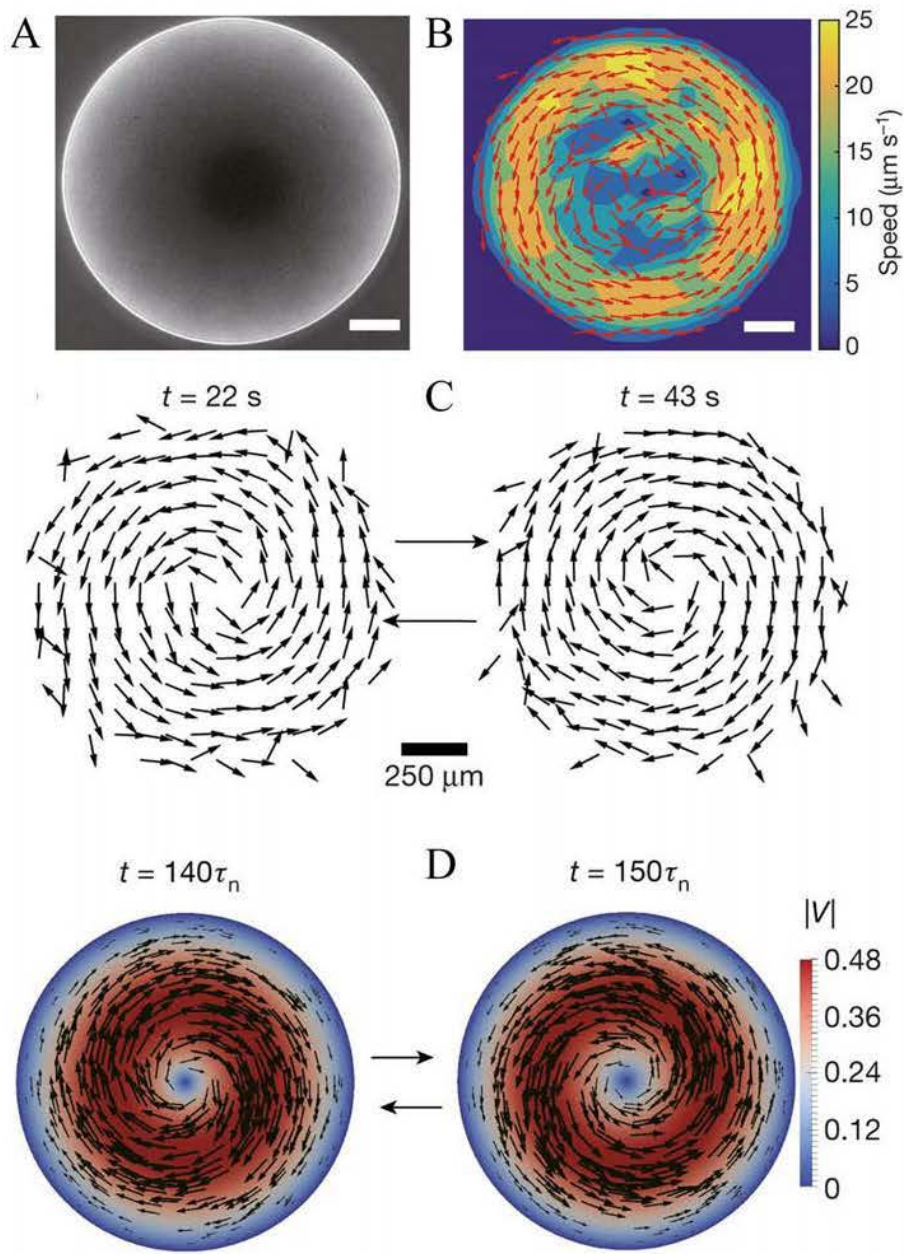

Figure 1.10: The dramatic effect of adding a small amount of DNA on the flow field of swimming bacterial suspensions. Fig. 1.10: (A) contrast image and (B) instantaneous velocity field of the unidirectional giant vortex observed when 200 ng/mL of DNA are added to a dense suspension of pushers. Fig. 1.10: (C) instantaneous velocity field and (D) velocity contour map of an oscillatory giant vortex obtained by increasing DNA concentration up to 800 ng/mL. Adapted from Ref.[96] with permission from [Springer Nature], Copyright [2021].

In light of these results based on E. T., it is natural to ask under which conditions the addition of polymers to a swarm of swimming bacteria can modify the observed turbulent-like flow structures. This question was addressed in the recent work of Liu *et al.*,[96] where the effect of addying DNA to a dense suspension of *E. Coli* was investigated by means of a combination of particle-image velocimetry (PIV) and rheometry. When the DNA concentration exceeded 50 ng/$\mu$L, the observed flow structure transitioned from a disordered ensemble of small vortices to a single, sample-spanning rotating vortex. With a further increase of the DNA concentration, the single vortex displays an oscillatory behaviour (see Fig. 1.9). The experimental data were also supported by simulations, which demonstrated that the observed transition from a rotating large vortex to an oscillating large vortex is driven by a balance between the active force exerted by the swimming bacteria and the viscoelasticity of the medium (see Fig. 1.10).

# 1.7  Perspectives

Over the past several decades, there has been an extraordinary progress in understanding the rheology of active fluids, arising mainly from theoretical research. Despite such promising experimental results, there is still a lack of rheological data concerning active particles in both Newtonian and non-Newtonian matrices. As discussed in Section 1.3, the viscosity measurements of pushers suspensions encounter key problems due to sensitivity limits of rotational rheometers. Microfluidic rheometers based on either flow visualisation[36,37] or on direct pressure sensing[84,101,102] have recently emerged as techniques that can successfully characterise the rheological properties of low viscosity fluids over a range of shear rates much larger than what can be explored by conventional techniques. It is envisaged that such techniques will be increasingly used to study the rheometric properties of active fluids.

It is worthy to highlight that examples of active fluids where the suspended particles are other than bacteria or synthetic swimmers have not yet been properly investigated. In particular, protein-based active fluids have not yet received much attention from the scientific community. Motor proteins are mechanochemical enzymes that can move along the cytoplasm of animal cells. Their activity is due to the hydrolysis of adenosine triphosphate (ATP).[103] Kinesins, dyneins are motor proteins that can move actively along microtubules, while and myosins can move along actin filaments. Microtubule motors can be plus-end motors and minus-end motors, depending on the "walk" direction along the microtubule cables within the cell. Mofrad *et al.*[104] highlighted the unique aspect of the cytoskeleton for which active and passive characteristics need to be considered at the same time. Moreover, cytoskeletal structures are highly stimuli-responsive, which further complicates experimental measurements aiming at establishing material properties.[105]

The recent observation concerning the onset of a non-linear flow for pusher suspensions in the superfluid-like regime represents an exciting finding which has stimulated a noteworthy research activity.[39,70,96] There is still a wide avenue for future work, specially focusing on the effect of polymer additives on the observed flow effects. The interaction between such collective bacterial motion and the well-known 'Elastic Turbulence' regime observed if a small amount of polymer is added to the buffer[99,100] deserves a detailed investigation. Moreover, the rheological properties of polymer solutions vary significantly if the concentration is changed across four concentration regimes (dilute, semi-dilute unentangled, semi-dilute entangled and concentrated[106]), and the effect of such variation on the bacterial motion should constitute the focus of a systematic study.